\documentstyle[12pt,aasms4,psfig]{article}

\slugcomment{to appear in the Astrophysical Journal 2001}  
 
\begin{document}
 
\title{A 250 GHz Survey of High Redshift QSOs from the Sloan Digital
Sky Survey}
 
\author{
C. L. Carilli$^{1}$,
F. Bertoldi$^2$,
M.P. Rupen$^1$,
Xiaohui Fan$^{3,5}$,
Michael A. Strauss$^3$,
K.M. Menten$^2$,
E. Kreysa$^2$,
Donald P. Schneider$^4$,
A. Bertarini$^2$,
M.S. Yun$^1$,
R. Zylka$^2$
}          

\affil{$^{1}$National Radio Astronomy Observatory, P.O. Box O,
Socorro, NM 87801, USA }

\affil{$^{2}$Max-Planck-Institut f\"{u}r Radioastronomie,
Auf dem H\"ugel 69, D-53121 Bonn, Germany}

\affil{$^{3}$Princeton University Observatory, Peyton Hall, Princeton,
NJ 08544, USA}

\affil{$^{4}$Dept. of Astronomy, Pennsylvania State University, 
University Park, PA 16802, USA} 

\affil{$^{5}$Institute for Advanced Study, Olden Lane, Princeton,
NJ 08540, USA}

\vskip 0.2in
\affil{ccarilli@nrao.edu}      

\begin{abstract}

We present observations at 250 GHz (1.2 mm), 43 GHz, and 1.4 GHz of a
sample of 41 QSOs at $z > 3.7$ found in the Sloan Digital Sky Survey.
We detect 16 sources with a 250 GHz flux density greater than $\rm 1.4~
mJy$.  The combination of centimeter and millimeter wavelength
observations indicates that the 250 GHz emission is most likely
thermal dust emission.  Assuming a dust temperature of 50 K, the
implied dust masses for the 16 detected sources are in the range $1.5
\rm ~ to ~ 5.9 \times 10^8$ M$_\odot$, and the dust emitting regions
are likely to be larger than 1 kpc in extent.  The radio-through-optical
spectral energy distributions for these sources are within the broad
range defined by lower redshift, lower optical luminosity QSOs. We
consider possible dust heating mechanisms, including UV emission from
the active nucleus (AGN) and a starburst concurrent with the AGN, with
implied star formation rates between 500 and 2000 M$_\odot$
year$^{-1}$.

\end{abstract}
 

\section {Introduction}

The existence of massive black holes at the centers of
galaxies has long been predicated on consideration of the energetics
of active galactic nuclei (AGN): accretion of matter onto a massive
black hole 
is an order of magnitude more efficient at converting mass into energy
than is stellar nucleosynthesis (Begelman, Blandford, \& Rees
1984).  The last few years have seen an
explosion of dynamical evidence for massive black holes at the center
of galaxies, and in the direct measurement of black hole masses from
the dynamics of circumnuclear regions
(Richstone et al. 1998; Miyoshi et al. 1996;
Ghez et al. 1998; Genzel et al.  2000; Tanaka et al. 1995).
These observations  indicate that the overwhelming majority of
spheroidal galaxies in the nearby universe contain massive black holes
(Kormendy 2000). Moreover, a clear correlation has been found between
the black hole mass and the velocity dispersion of the stars in the
spheroid (Gebhardt et al. 2000, Ferrarese \& Merritt 2000). This
remarkably tight correlation suggests a `causal
connection between the formation and evolution of the black hole and
the bulge' (Gebhardt et al. 2000; Kormendy 2000;
Richstone et al. 1998). 

Support for this general idea comes from: (i)  the observed space
density of high redshift Quasi-Stellar Objects (QSOs), which would require 
a massive black hole in most low $z$ spheroidal galaxies
(Kormendy \& Ho 2000), (ii) the observation
of similar rapid increases in the space densities of AGN and starburst
galaxies from $z = 0$ to $z = 2$ (Boyle \& Terlevich 1998, Blain et
al. 1999), and (iii) the observation of co-eval AGN and star formation
activity in some nearby nuclear starburst galaxies (Rigopoulou et
al. 1999; Carilli, Wrobel, \& Ulvestad 1999). 
On the other hand, the QSO population shows an abrupt turn-over in 
the  co-moving number density  at $z \ge 3$ 
(Schmidt, Schneider, \& Gunn 1995a; 
Kennefick, Djogovski, \& de Carvahlo 1995;
Fan et al. 2001b). 
It is unclear  whether such a turnover exists in 
the starburst population (Blain et al. 1999, Dunlop
2000).  Also, questions remain about the 
relative timescales for the commencement and duration of the starburst
and AGN processes (Sanders \& Mirabel 1996), and the origin
at very high redshift of the `seed' black hole (Richstone et
al. 1998). 

A particularly intriguing observation that has fueled the debate over
co-eval starbursts and AGN at high redshift is the recent
detection at 250 GHz of copious thermal dust emission 
from high redshift QSOs by
Omont et al. (1996a).  
Omont et al. observed a sample of $z > 4$ QSOs from the Automatic
Plate Measuring (APM) survey and found that 6 of 16 sources show dust
emission with 250 GHz flux densities of 3 mJy or greater, with implied
Far Infrared (FIR) 
luminosities, $\rm L_{FIR} > 10^{12}$ L$_\odot$, and dust masses $\ge
10^8$ M$_\odot$.  Follow-up observations of three of these dust-emitting
QSOs revealed CO emission as well, with implied molecular gas masses
$\approx$ few $10^{10}$ M$_\odot$ (Guilloteau et al. 1997, 1999; Ohta
et al. 1996; Omont et al. 1996b; Carilli, Menten, \& Yun 1999).

Two different mechanisms have been proposed for heating the dust
in high redshift QSOs.  Given the large dust and gas
masses, Omont et al. (1996a) make the circumstantial argument that
star formation is inevitable, and hence that the dominant dust heating
mechanism may be star formation.  Supporting evidence comes from deep
radio observations at 1.4 GHz, which show that the ratio of the radio
continuum to submillimeter continuum luminosity from some of these
sources is consistent with the well established radio-to-far IR
correlation for low redshift star forming galaxies (Yun et al. 2000).
And Omont et al. (2001) make the
simple point that copious star formation is required in order to
produce the dust.
The implied star formation rates are so extreme, $\ge$ 10$^3$ M$_\odot$
year$^{-1}$, that a significant fraction of the stars in the QSO
host galaxy can be formed in $< 10^9$ years.

Alternatively, the  dust may be heated by
radiation from the AGN.  Sanders et al. (1989) argue that 
radiation from the AGN is
the dominant dust heating mechanism, since in most cases it requires
the absorption of only a small fraction ($\le 20\%$) of the AGN UV
luminosity by dust, even for the high redshift sources (Carilli et
al. 2000). An important 
constraint on AGN dust heating models is provide by the 
minimum size of about 1 kpc for the dust emission region one derives
from the observed luminosity and dust temperature
(see section 3).  Hence, while the optical `big blue
bump' emission is thought to arise in a hot ($10^{4.5}$ K) accretion
disk on pc-scales, the rest-frame far IR emission must be due to warm
dust on kpc-scales.  In the Sanders et al. (1989) model for
AGN-powered IR emission from QSOs, dust heating on kpc scales is
accomplished by assuming that the dust is distributed in a kpc-scale
warped disk, thereby allowing UV radiation from the AGN to illuminate
the outer regions of the disk. An alternate mechanism for large-scale
dust heating has been proposed by Maloney, Hollenbach,
\& Tielens (1996) in which hard X-rays penetrate the very high column
density gas in the vicinity of the AGN.

In this paper we present extensive observations at cm and
mm wavelengths of a sample of high redshift QSOs from the
Sloan Digital Sky Survey (SDSS; York et al. 2000). 
Our observations include sensitive radio
continuum imaging at 1.4 and 43 GHz with the Very Large Array (VLA),
and photometry at 250 GHz using the Max-Planck Millimeter Bolometer
array (MAMBO) at the IRAM 30m telescope.  These observations are a
factor three more sensitive than previous studies of high redshift
QSOs at these frequencies (Schmidt et al. 1995b; Omont et al. 1996a;
Stern et al. 2000). Details of the multifrequency 
radio continuum imaging will be given elsewhere (Rupen et
al. 2001, in preparation). In this paper we present the radio results
that are relevant to the analysis and interpretation 
of the dust continuum emission.
We compare the dust emission properties of these sources with
their optical and radio continuum and spectral properties, 
and with those of lower redshift sources. 
We assume $\rm H_o = 50$ km s$^{-1}$ Mpc$^{-1}$, $\rm q_o = 0.5$, 
$\Lambda = 0$, and
we define the spectral index, $\alpha$, as a function of
frequency, $\nu$, as: $\rm S_\nu \propto \nu^\alpha$. 

\section{The Sample}

The QSO sample is the result of optical spectroscopy of objects of unusual
color from the northern Galactic Cap and the Southern Equatorial
Stripe of the SDSS. The survey has yielded over 100 QSOs with $z \ge 3.6$,
including eight of the ten highest redshift QSOs known (Fan et
al. 1999, 2000b, 2001a,b; Schneider et al. 2000a,b, 2001;
Zheng et al. 2000).

We observed 41 of these high redshift QSOs, taken from Fan et
al. (1999), Fan et al. (2000a), and Schneider et al. (2000a).
The QSO properties are listed in Table 1.
The observed sources span a range of $\rm M_B = -26.1 \rm ~to~ -28.8$,
and a redshift range of $z =$ 3.6 to 5.0.  Comparative numbers for
the APM sample of 16 QSOs
observed by Omont et al. (1996a) are $\rm M_B = -26.8
\rm ~to~ -28.5$, and $z = $ 4.0 to  4.7.

The sources in Table 1 have a mean optical spectral power law
index, $\alpha = -0.67$, with a dispersion of $0.29$,  
a mean readhift, $z = 4.20$, with a dispersion of $0.44$, 
and a mean absolute blue magnitude, $\rm M_B =  -27.07$, with 
a dispersion of $0.57$.
We can compare these values to those of the
39 QSOs at $z > 3.6$  from the 
SDSS Fall Equatorial stripe presented in 
Fan et al. (2001b). The Fall Equatorial stripe sample is 
a statistically complete sample of high-$z$ 
QSOs selected according to the same criteria as the sources listed in
Table 1. The Fall Equatorial stripe sample has
a mean optical spectral power law index, $\alpha = -0.79$,
with a dispersion of $0.34$,  
a mean readhift, $z = 4.06$, with a dispersion of $0.39$, 
and a mean absolute blue magnitude, $\rm M_B =  -26.82$, with 
a dispersion of $0.58$.
Hence, within the scatter of the distributions,
the sources in Table 1 can be considered representative of 
the complete high-$z$ QSO sample from the Fall Equatorial stripe.
 
\section{Observations}

Observations were made using MAMBO (Kreysa et al. 1999) at the IRAM 
30m telescope on Pico Veleta in Spain, 
in December 1999 and February 2000. MAMBO is a 37
element bolometer array sensitive between 190 and 315 GHz.  The
half-power sensitivity range is 210 to 290 GHz, but
the overall profile is asymmetric, with a sharp rise in
sensitivity at low frequency, then a gradual  decrease
in sensitivity to higher frequency. Convolving the frequency response
of the bolometer array with the atmospheric transmission curve
(which also decreases with increasing frequency), and with the 
rising spectrum of a typical dust emitting source at 
high redshift, leads to an  effective central frequency of 250 GHz.  
The beam for the feed horn of each bolometer is matched
to the telescope beam of 10.6$''$, and the bolometers are arranged in
a hexagonal pattern with a beam separation of 22$''$. Observations
were made in standard on-off mode, with 2 Hz chopping of the secondary
by 50$''$ in azimuth.  The data were reduced using
the  MOPSI software package of the
Max-Planck-Institut f\"{u}r Radioastronomie (Zylka 1998).  Pointing was
monitored every
hour, and was found to be repeatable to within 2$''$. The sky opacity
was monitored every hour; zenith optical depths ranged between 0.23 and
0.36. Gain calibration was performed using observations of Mars,
Uranus, and Ceres.  We estimate a 20$\%$ uncertainty in absolute
flux density calibration based on these observations.  The target
sources were centered on the central bolometer in the array,
and the temporally correlated variations of the sky signal
(sky noise) detected in the remaining bolometers were subtracted from
all the bolometer signals. The total on-target plus off-target
observing times for the sources are listed in 
Table 1. The typical rms sensitivity in 1 hour was 0.5 mJy for the
on-source bolometer beam. 

The 1.4 GHz VLA observations were made in the A (30 km) and BnA (mixed
30 km and 10 km) configurations on July 8, August 14, September 30,
and October 8, 1999, using a total bandwidth of 100 MHz with two
orthogonal polarizations.  Each source was observed for a total of 2
hours, with short scans made over a large range in hour angle to
improve Fourier spacing coverage. Standard phase and amplitude
calibration were applied, as well as self-calibration using background
sources in the telescope beam. The absolute flux density scale was set
using observations of 3C 48.  The final images were generated using
the wide field imaging and deconvolution capabilities of the AIPS task
IMAGR.  The theoretical rms noise value is between 20 and 30 $\mu$Jy
beam$^{-1}$, depending on the range of telescope elevations over which
the source was observed,  and for about
half the sources the measured noise values are in this range. For the 
remaining sources, the noise levels are 
significantly higher due to side-lobe confusion by bright sources.
The Gaussian restoring CLEAN beams were between 1.5$''$ and 3$''$
FWHM.  

Some of the sources had previous VLA detections as part of the FIRST
survey (SDSS J1053-0016, SDSS J1235-0003, SDSS J1412-0101; Becker,
White, \& Helfand 1995).  We re-observed these sources, and obtained
the same flux densities to within 15$\%$ of the FIRST values 
in all cases.

We chose four of the dust emitting sources, (SDSS J0150+0041, SDSS
J0255+0048, SDSS J0338+0021, SDSS J1112+0050) as a
sub-sample for 43 GHz observations, in order to search for high
frequency, flat spectrum emission that might be synchrotron
self-absorbed at 1.4 GHz. Observations were made on
Feb 20, 2000, using the VLA in the B (10 km) configuration in standard
continuum mode.  Fast switching phase calibration was employed
(Carilli \& Holdaway 1999), and observing times were 0.5 hr at 43 GHz,
resulting in an rms noise of 0.35 mJy beam$^{-1}$.

\section{Results and Analysis}

The results of our 250 GHz survey of high redshift QSOs are given in
Table 1.  These include: 
(i) flux densities at 250 GHz (S$_{250}$; column 3),
(ii) flux densities at 1450 ${\rm \AA}$ (S$_{1450}$; column 4),
(iii) flux densities at 1.4 GHz at the position of the optical QSO
(S$_{1.4}$; column 5),
(iv) absolute blue magnitudes (column 6; Fan et al. 1999, 2000a),
(v) observing time at 250 GHz, and (vi) notes on source
properties (column 8).  
In the following analysis we consider a source 
with a measured flux density three times greater
than the rms noise (3$\sigma$) to be a valid detection. 
Such sources are marked with an asterisk in Table 1.
We should also point out that the mm, cm, and optical
observations were not simultaneous, and hence that 
variability could affect the observed spectra at some level.


We detect 16 of the 41 sources at S$_{250} \ge$ 1.4 mJy.
Only four of the sources have  S$_{250} \ge 3$ mJy.
A higher detection rate
at  the limit of S$_{250} \ge 3$ mJy was found in 
the survey of the APM QSO sample by 
Omont et al. (1996), in which they  detected 6 of 16 sources.
This difference in detection rates appears to be statistically
significant: if there were a true universal fraction of 
10$\%$ of high-$z$ QSOs with S$_{250} \ge 3$ mJy, then there is
about a 1$\%$ chance that in 16 objects 6 or more would meet this
criterion. The optical/near IR color selection criteria for the APM
sample are similar to those of the SDSS sample. 
The one significant difference is that the  
rest frame blue luminosities for the APM are somewhat higher
on average (by about 0.5 mag), than those of the SDSS sources in Table
1 (see section 2). However, 
given the lack of a strong correlation between M$_{\rm B}$ and
S$_{250}$ (see Figure 2 below), the cause for the
possibly higher detection rate
at high S$_{250}$ for the APM sample relative to the SDSS sample
remains unknown. Further observations of the mm properties of
high $z$ QSOs that are in progress may clarify this issue 
(Omont et al. 2001). 

Considering the radio properties of the sources in Table 1, nine
sources in Table 1 are detected at 1.4 GHz at $\ge 3\sigma$.  For the
nine radio-detected sources in Table 1, eight are unresolved, with
size limits (FWHM) of about 1$''$, while SDSS J0232-0000 is possibly
extended, with a size $\sim 1.3''$.  Miller et al. (1990) suggested a
division between radio loud and radio quiet QSOs at a rest frame 5 GHz
spectral luminosity of $10^{33}$ erg s$^{-1}$ Hz$^{-1}$. This
corresponds to a flux density of S$_{1.4}$ = 1 mJy for a source at $z
= 4.2$ assuming a radio spectral index of --0.8.  According to this
criterion, 2 of 16 SDSS QSOs detected at 250 GHz can be considered
radio-loud (SDSS J1235--0003 and SDSS J1412--0101), while 2 of 25
sources not detected at 250 GHz can be considered radio-loud (SDSS
J0153--0011 and SDSS J1053--0016).  Both fractions are consistent with
the value of 10$\%$ of optically selected high redshift QSOs being
radio loud according to this criterion (Schmidt et al. 1995b; Stern et
al. 2000).

\subsection{Evidence for thermal dust emission at 250 GHz}

We first address the question of whether the 250 GHz emission 
is thermal dust emission, or non-thermal synchrotron radiation.
For most of the sources detected at 250 GHz, 
the 1.4 GHz flux density (or upper limit) is an
order of magnitude or more  below the 250 GHz flux density. 
Two of the sources have 1.4 GHz flux densities 
comparable to, or larger than,  the 250 GHz flux density.
The source SDSS J1412--0101 has S$_{1.4} = 3.9$ mJy and S$_{250} =
4.5$ mJy. Observations of this source at  8.4 GHz revealed
a 1.1 mJy source (Rupen et al. 2001), 
such that the  cm source has a falling spectrum of 
index --0.7, making it unlikely that the mm continuum is a
continuation of the synchrotron emission spectrum from the AGN.
The source SDSS J1235--0003 has S$_{250} = 1.6$ mJy, 
and S$_{1.4} = 4.5$ mJy. Observations 
on the same day as those at 1.4 GHz found
a 5 GHz flux density of S$_{5} = 17$ mJy, and 
a 15 GHz flux density of S$_{15} = 7$ mJy. The source was
unresolved at all frequencies. In this case, it remains possible that
the 250 GHz emission is non-thermal. 

It is possible that emission 
from a compact AGN could be synchrotron self-absorbed at low
frequency. To check this idea, we performed 43 GHz observations of 
4 of the 250 GHz detected sources, as listed in section 3; 
none of the sources were detected.
Assuming a 2$\sigma$ upper limit of 0.7 mJy at 43 GHz implies
a rising spectrum between 43 and 250 GHz of 
index $\alpha >0.5$.
While this limit is well below the index 
expected for thermal dust emission ($\alpha_{43}^{250} \approx
3$ for a source at $z = 4.2$), it does argue against 
compact, synchrotron self-absorbed sources
as the origin of the 250 GHz emission, as such
sources typically show mm spectral indices  
$\alpha_{\rm mm} \sim 0 \pm 0.3$ (Sanders et al. 1989). In the
following discussion we 
will  assume that the 250 GHz emission from the sources in Table 1
is thermal dust emission, with the possible exception of
SDSS J1235--0003. These data imply that
in mm surveys of high redshift QSOs, the fraction
of flat spectrum, radio loud AGN with synchrotron emission extending
into  the mm-regime  is at most a few percent. 
Studies of lower redshift QSOs show that the fraction of 
flat spectrum  radio loud sources in optically
selected samples is also  only a few percent (Hopkins et
al. 1999).

A source with S$_{250} = 1$~mJy at $z = 4.2$ and 
a dust spectrum of the type seen in the low redshift starburst
galaxy Arp 220 (corresponding roughly to a modified black body
spectrum of dust emissivity index = 1 and temperature = 50 K) has 
an FIR luminosity, L$_{\rm FIR}$,  
of $1.1 \times 10^{12}$ L$_\odot$,
where $\rm L_{FIR}$ corresponds to the integrated luminosity between
42 and 122 $\mu$m (Condon 1992). 
If we assume that star-formation is the mechanism giving rise to
the dust emission, we can use the relations in  Omont et
al. (2001) and Carilli et al. (2001)
to derive the dust mass, $\rm M_D$, and total star formation rate, SFR,
from  $\rm S_{250}$. The relations are:~
$\rm M_D = 1.1\times 10^8\times S_{250}$ M$_\odot$,
and $\rm SFR = 360\times S_{250} ~ M_\odot~ year^{-1}$,
with  S$_{250}$ in mJy. This again assumes $\rm T_D = 50$ K. 

A conservative lower  limit to the solid angle,  $\Omega_s$, 
of a dust emitting source
can be derived from the observed flux density
making the extreme assumption of optically thick emission
(Downes et al. 1999). 
Assuming $\rm T_D \sim 50~ K$ (Benford et al. 1999) and  $z \sim 4$,   
leads to:
$ \rm \Omega_s > 0.0043~ S_{250} ~ arcsec^2, $
where S$_{250}$ is the 250 GHz flux density in mJy.
This implies absolute lower limits to the angular diameters
of the sources in Table 1 between $0.08''$ and $0.15''$, corresponding
to physical sizes of 0.5 to  1 kpc.
Given the likelihood that the observed 250 GHz emission
is from an optically thin grey body, it is almost certain
that the sources are larger than these lower limits.
Also,  this size corresponds to the total emitting area.
Multiple smaller regions  within the 10.6$''$ telescope beam
are certainly possible (see section 5). 
Observations at mm wavelengths with sub-arcsecond resolution 
are required to constrain the source sizes. 

\subsection{Trends with Redshift and M$_{\rm B}$}

Figure 1 shows values of S$_{250}$ versus redshift for the 
41 QSOs in Table 1. The solid line is the expected flux density
of Arp 220. This curve shows clearly the effect of the large `inverse
$K$'  correction for thermal dust emission at mm and submm
wavelengths, with the observed flux density of a source such as Arp
220 remaining roughly constant over the observed redshift range
(Blain and Longair 1993). 
There is an interesting 
trend for a lower detection rate for the SDSS
QSOs with increasing redshift: for $z < 4.4$ we detect 14 of 28
sources while for $z > 4.4$ we detect 2 of 13 sources. 
A larger sample of sources, over a larger redshift range, is
required to verify this trend.

Figure 2 shows values of L$_{\rm FIR}$ versus M$_{\rm B}$ for
the 41 QSOs  in Table 1.
At the limit of  L$_{\rm FIR} > 1.8\times10^{12}$ L$_\odot$,
we detect 9 of 23 sources
with M$_{\rm B} > -27$, and 7 of 17 sources with
M$_{\rm B} < -27$. This is consistent
with the lack of a strong correlation between 
M$_{\rm B}$ and  L$_{\rm FIR}$ in the lower redshift, lower optical 
luminosity sample of Palomar-Green (PG)
QSOs presented  by Sanders et al. (1989; see also Chini, Kreysa, \&
Biermann 1989), and is indicative of the large scatter in
the dust properties of QSOs in general (see Figure 4 below),
and the relatively narrow range in M$_{\rm B}$ in Figure 2. 

\subsection{Comparison to the Radio-to-FIR Correlation for Star
Forming Galaxies}

Figure 3 shows the relationship between redshift and 
the 250-to-1.4 GHz spectral
index for a star forming galaxy taken from 
the model of Carilli \& Yun (2000) based on 17 low redshift galaxies
(roughly equivalent to a modified black body spectrum 
with $\rm T_D = 50 K$ and dust emissivity index = 1.5). 
The relationship relies on the tight radio-to-FIR correlation seen
for star forming galaxies in the nearby universe (Condon 1992).
The dotted curve gives the rms scatter for the 17 galaxies.
The solid symbols are the results for the SDSS QSOs detected at
250 GHz, while the open symbols are for the APM QSO sample from 
Omont et al. (1996) and Yun et al. (2000). 
The arrows  indicate non-detections in the
radio, and hence lower limits to the spectral index.  On this diagram, 
a point located below the curve would indicate  a
source that is radio-loud relative to the 
standard radio-to-FIR correlation for star forming galaxies, 
while  a point located above the curve 
would indicate a source which is radio-quiet relative to this
relationship (Condon 1992). The lower limits are below the curve, 
and hence are consistent with a star forming galaxy spectrum,
although a factor two better sensitivity at 1.4 GHz is required to
provide more stringent constraints in this regard. 

The radio-detected QSOs fall below the star forming galaxy
curve in Figure 3.  Two of the
sources (SDSS J1235--0003 and SDSS J1412--0101) are clearly radio loud
AGN (see section 4.0). The fainter radio detections may indicate
radio emission from the AGN, but at a level below that
required for the source to be defined as 
radio loud according to  Miller et al. (1990). 
An alternative explanation for
these sources is a warmer dust component: increasing the dust
temperature by a factor two or so could give rise to the observed
offset with respect to the model in Figure 3 (Blain 1999).

\subsection{Spectral Energy Distributions}

Figure 4 shows the mean radio-through-optical Spectral Energy
Distributions (SEDs) for 
radio-loud and radio-quiet QSOs derived from observations of a large
sample of PG QSOs by Sanders et al. (1989). 
The solid curve shows the SED for radio loud QSOs, while the 
dashed curve shows the SED for radio quiet QSOs. The hatched regions
indicate the scatter in the measured values for the PG sample. 
Note that the QSOs in the PG sample are typically an
order of magnitude less
luminous in the rest-frame UV than the high redshift sources,
with most of the PG sources in the range: $\rm M_B = -23~ to~ -27$.

The data points in Figure 4 show the results for the mm detections
in the SDSS and APM QSO samples, normalized  by their blue spectral
luminosities. The normalized mm detections all fall within
the range defined by the PG sample. The upper limits for
the non-detected sources at  250 GHz would fall at the low end of the
range defined by the Sanders et al. (1989) SEDs.  

The two radio loud SDSS QSOs according to the Miller et al. 
(1990) definition detected at 250 GHz are clearly evident 
in the normalized SEDs in Figure 4,
while the normalized
upper limits for the non-detected sources at 1.4 GHz 
are consistent with the Sanders et al. (1989) radio quiet QSO SED. 
The remaining three radio-detected SDSS QSOs in Figure 4
fall in-between 
the radio quiet and radio loud SEDs, as do the QSOs from
the APM sample. 
The large gap between the radio loud and radio quiet
SEDs from Sanders et al. (1989), as reproduced in Figure 4, 
represents the possible bi-modality
of the radio properties of QSOs, as suggested by Stocke et
al. (1992). The reality of this bi-modality  has been
called into question recently by White et al. (2000), and
Stocke et al. (1992) show that the bi-modality
is less clearly delineated
for high blue luminosity QSOs relative to lower luminosity sources.
Further analysis 
of the radio detections within this sample,   including
studies of radio spectra and spatial structure, 
will be given in Rupen et al. (2001). 

\subsection{Broad Absorption Lines}

Omont et al. (1996a) suggest a possible correlation between the
Broad Absorption Line (BAL) phenomenon and thermal dust emission
in QSOs. In the APM sample of Omont et al. (1996a), 
two of six sources detected at 250 GHz have BAL features.
The BAL fraction for high $z$ QSOs in general is about 10$\%$
(Fan et al. 2001a). 

Our data on the SDSS sample provide only marginal support for this
idea, with 4 of the 16 sources detected at S$_{250}$ GHz
showing evidence for BAL features, as compared to 3 sources with
BALs out of the 25 non-detected sources at 250 GHz. 

\section{Discussion}

We detect 16 of 41 high redshift QSOs from the SDSS survey
with S$_{250} \ge 1.4$ mJy. 
Assuming $\rm T_D = 50~ K$,
we show that the 250 GHz emission is most likely
thermal emission from warm dust on a scale $> 1$ kpc in the host
galaxy of the QSO.  The implied dust masses for the sources are in the
range of $1.5 \rm ~ to ~ 5.9 \times 10^8$ M$_\odot$.  
If the dust is heated by star formation, the star formation
rates are in the range 500 to 2000 M$_\odot$ year$^{-1}$.  Sensitive
radio observations show a radio-loud fraction (according to the
definition of Miller et al. 1990) for the sample as a whole of 10$\%$,
consistent with previous observations of high redshift QSOs, with no
clear relationship between dust emission and radio-loud QSOs.

The blue-normalized optical-through-radio SEDs for these sources are
within the broad range delineated by the PG sample of lower redshift,
lower luminosity QSOs (Figure 4).  This result suggests that for the
sources detected at 250 GHz, we are not seeing a major new broad-band
spectral emission `feature' in high redshift QSOs relative to the SEDs
of lower redshift QSOs.  Conversely, the blue-normalized limits for
the non-detections at 250 GHz in the SDSS high $z$ QSO sample are at
the low end of the distribution defined by low $z$ QSOs, leaving open
the possibility of a population of high $z$ sources which are
under-luminous in the rest frame submm relative to the SED of low $z$
QSOs.  While this result could be used to argue against models
involving a major burst of star formation in the host galaxies of high
redshift QSOs, the large scatter in the mm SEDs could effectively mask
such a phenomenon.  Also, Sanders et al. (1989) point out that from cm
to submm wavelengths, the radio quiet SED model based on the PG QSO
sample is consistent with the observed SEDs for star forming galaxies.


A number of studies have addressed the question of the dust heating
mechanism in high $z$ AGN by SED modeling, with mixed results.
Radiative transfer models by Rowan-Robinson (1999) suggest that the
observed SEDs at rest frame wavelengths $> 50 \mu$m are consistent
with an extreme starburst model for most high redshift dust emitting
QSOs. On the other hand, models by Andreani et al. (1999) and Willott
et al. (1999) suggest that the dust emission spectra from 3$\mu$m to
30$\mu$m can be explained by dust heated by the AGN.  The cm-to-mm
SEDs may also indicate star formation in a few sources (Yun et
al. 2000), however the radio limits in most sources are inadequate to
constrain such a model (Figure 3).  Mid-infrared spectroscopy 
using  the Space Infrared Telescope Facility of the
PAH features and other spectral lines may provide a star formation vs. AGN
diagnostic (Genzel et al. 1998).

Spatially resolved observations may be critical in addressing this
issue, since a dust emitting region spatially distinct from the
optical QSO can be used to argue for at least some dust heating by
star formation. Such an argument has been used convincingly in the
case of CO and dust emission in the vicinity of a number of high
redshift radio galaxies and AGN (Papadopoulos et al. 2000, 2001;
Ivison et al. 2000). To date, the only high redshift QSO in which such
a spatial segregation has been observed is 
the $z = 4.7$ QSO BR1202-0725, in which the
cm and mm continuum, and the CO line emission, show two sources of
roughly equal strength separated by 4$''$, while the optical line and
continuum emission is restricted to a single source (Hu, McMahon, \&
Egami 1996; Omont et al. 1996b; Ohta et al. 1996; Yun et al. 2000;
Kohno et al. 2000, in preparation).

The question of the dominant dust heating mechanism, ie. AGN versus
star formation, is difficult to answer for low redshift sources
(Rigopoulou et al. 1999).  Addressing the question at high redshift is
just that much more difficult. The sources detected at 250 GHz in
Table 1 present an ideal opportunity for addressing this issue in
detail.  A number of follow-up observations of these sources are
required to properly address this question, including: (i) more
sensitive radio continuum observations to better characterize the cm-to-mm
SEDs, (ii) submm observations to determine the dust temperature, (iii)
high resolution (sub)mm observations to determine the spatial distribution
of the dust, and (iv) CO observations to better constrain the gas
mass and kinematics.  Such observations are at the limit of what is
possible with current instrumentation, but will become routine with
the expanded VLA and the Atacama Large Millimeter Array.

\vskip 0.2truein 

The VLA is a facility of the National Radio
Astronomy Observatory (NRAO), which is operated by Associated
Universities, Inc. under a cooperative agreement with the National
Science Foundation.
This work was based on observations carried out with the IRAM 30 m
telescope.  IRAM is supported by INSU/CNRS (France), MPG (Germany) and
IGN (Spain).  This research made use of the NASA/IPAC Extragalactic
Data Base (NED) which is operated by the Jet Propulsion Lab, Caltech,
under contract with NASA.  CC acknowledges support from the Alexander
von Humboldt Society.  DPS acknowledges support from National Science
Foundation Grant AST99-00703.  XF and MAS acknowledge support from the
Research Corporation, NSF grants AST96-16901 and AST00-71091
and an Advisory Council Scholarship.


\newpage

\newpage

\begin{table}
\caption{SDSS High Redshift QSOs} 
\vskip 0.2in
\begin{tabular}{cccccccc}
\hline
\hline
Source & $z$ & S$_{250~ \rm GHz}$ & S$_{1450~ {\rm \AA}}$ & 
S$_{1.4~ \rm GHz}$ & M$_{\rm B}$ & Obs. time & Notes \\
SDSS & ~  & mJy & mJy & mJy & ~ & sec & ~ \\
\hline
J003525.29+004002.8 & 4.75 & 0.3$\pm$0.5 & 0.038 &  0.024$\pm$0.027 &
--26.66 & 3500 & ~ \\
J010619.25+004823.4 & 4.43 & --0.4$\pm$0.4 & 0.121 & 0.601$\pm$0.043$^{*,a}$ &
--27.70 & 4640 & DPOSS/APM \\
J012403.78+004432.7 &  3.81 &  2.0$\pm$0.3$^*$ & 0.215 &   0.11$\pm$0.048
& -28.19 & 5740 & ~ \\
J012650.77+011611.8 & 3.66 & --0.6$\pm$0.6  & 0.056 & 0.21$\pm$0.023$^*$ &
--26.62 & 2180 & ~ \\
J015048.83+004126.2 & 3.67 &  2.2$\pm$0.4$^*$ & 0.160 &  --0.004$\pm$0.019
& -27.75 & 4590 &  BAL \\
J015339.61--001104.9 & 4.20 & 0.9$\pm$0.5 & 0.100 &  5.3$\pm$0.041$^*$ &
--27.27 & 3210 &  ~ \\
J021102.72--000910.3 & 4.90 & 0.5$\pm$0.3 & 0.039  & 0.033$\pm$0.030 &
--26.63 & 7650 & ~ \\
J023231.40--000010.7 & 3.81 &  1.8$\pm$0.3$^*$ & 0.043 &  0.066$\pm$0.022$^*$
& -26.54 & 7350 & ~ \\
J025112.44--005208.2 & 3.78 & 2.4$\pm$0.6$^*$ & 0.058 & -0.03$\pm$0.024 &
-26.74 & 2170 & ~ \\ 
J025518.58+004847.6 & 3.97 &  2.1$\pm$0.4$^*$ & 0.125  &  0.043$\pm$0.031
& -27.67 &  3350 &  BAL \\
J031036.97--001457.0 & 4.63 &  -0.5$\pm$0.4 & 0.035 &  0.068$\pm$0.025
& --26.78 & 4490 &  ~ \\
J032608.12--003340.2 & 4.16 & 1.5$\pm$0.4$^*$ & 0.080 &  0.068$\pm$0.025 &
--27.21 & 4770 &  DPOSS/APM \\
J033829.31+002156.3 & 5.00 & 3.7$\pm$0.3$^*$ & 0.047 &   0.037$\pm$0.025 &
-26.56 & 8000 & BAL$^b$ \\
J105320.43--001649.3 & 4.29 & 0.1$\pm$0.2 & 0.063  &  11.5$\pm$1.0$^*$ &
--27.04 &  10360 & DPOSS/APM \\
J111246.30+004957.5 & 3.92 &  2.7$\pm$0.5$^*$ & 0.091 &  0.364$\pm$0.050$^*$ &
--27.50 & 2830 &  ~ \\
J111401.48--005321.1 & 4.58 &  0.2$\pm$0.3 & 0.047 &
--0.036$\pm$0.028 & --26.82 & 6880 & ~ \\
J112253.51+005329.8 & 4.57 &  --0.1$\pm$0.3 & 0.076 & 0.096$\pm$0.040
&  --27.35 & 10030 & ~ \\
J120441.73--002149.6 & 5.03 &  0.6$\pm$0.4 & 0.087 &  0.064$\pm$0.046
& --27.64 & 4300 &  \\
J122600.68+005923.6 & 4.25 & 1.4$\pm$0.4$^*$ &  0.086  &  0.016$\pm$0.027
& --27.36 & 4850 & ~ \\
J123503.04--000331.8 & 4.69 &  1.6$\pm$0.4$^*$ & 0.028    &  18.8$\pm$0.4$^*$
& --26.29 &  6320 &  ~ \\
\hline
\end{tabular}
~$^a$An astrisk indicates a 3$\sigma$ detection.
~$^b$Songaila et al. 1999
\end{table}

\clearpage
\newpage

\begin{table}
\vskip 0.2in
\begin{tabular}{cccccccc}
\hline
\hline
Source & $z$ & S$_{250~ \rm GHz}$ & S$_{1450~ {\rm \AA}}$ & 
S$_{1.4~ \rm GHz}$ & M$_{\rm B}$ & Obs. time & Notes \\
SDSS & ~  & mJy & mJy & mJy & ~ & sec & ~ \\
\hline
J131052.52--005533.4 & 4.14 & --0.3$\pm$0.5 &  0.097  &
--0.11$\pm$0.22 & --27.46 & 2380 & ~ \\
J132110.82+003821.7 & 4.70 &  0.4$\pm$0.4 & 0.033   &  0.036$\pm$0.030
& --26.47 & 4425 & ~ \\
J140554.07--000037.0 & 3.55 &  1.0$\pm$0.4 & 0.120    &
0.005$\pm$0.047 & --27.40 & 4540 & BAL \\
J141205.78--010152.6 & 3.73 &  4.5$\pm$0.7$^*$ & 0.068   &  3.9$\pm$0.2$^*$ &
--26.91 & 2160 & BAL \\ 
J141315.36+000032.1 & 4.08 &  2.0$\pm$0.8 & 0.034    &
0.040$\pm$0.030 & --26.30 & 1664 & ~ \\
J141332.35--004909.7 & 4.14 &  2.5$\pm$0.5$^*$ &  0.070 &  0.009$\pm$0.027
& --27.10 & 3730 &  BAL \\
J142329.98+004138.4 &  3.76 & 0.4$\pm$0.8 & 0.046 & 0.003$\pm$0.028 & --26.49 &
1270 & ~ \\
J142647.82+002740.4 & 3.69 &  3.9$\pm$0.8$^*$ & 0.056 & 0.055$\pm$0.033 &
--26.69 & 1290 & ~ \\
J144428.67--012344.1 & 4.16 &  --0.4$\pm$0.7 & 0.052   &
--0.003$\pm$0.027 & --26.80 & 2060  & ~ \\
J144758.46--005055.4 & 3.80 & 5.4$\pm$0.8$^*$ &  0.051   &
0.050$\pm$0.024  & --26.62 & 1160 & ~ \\
J151618.44--000544.3 & 3.70 &  --0.1$\pm$0.6 & 0.033   &
0.050$\pm$0.045 & --26.11 & 3220 &  ~ \\
J152740.52--010602.6 & 4.41 & --0.1$\pm$0.5 & 0.052   &
0.005$\pm$0.025 & --26.87 & 3690 & ~ \\
J153259.96--003944.1 & 4.62 & 0.2$\pm$0.7 & 0.063 & $<0.06^c$ &  --27.16 &
1418 & ~ \\
J160501.21--011220.0 & 4.92 & 0.1$\pm$0.4 & 0.061   & $<0.06^c$ & --27.23 &
4590 &  BAL \\
J161926.87--011825.2 &  3.84 & 2.3$\pm$0.6$^*$ & 0.047 & 0.047$\pm$0.025 &
--26.54 & 2560 & ~ \\
J162116.91--004251.1 & 3.70 &  0.1$\pm$0.3 & 0.394   &
0.019$\pm$0.027 & --28.81 &  5990 & ~ \\ 
J165527.61--000619.2 & 3.99 & --0.1$\pm$0.4 & 0.036 & 0.010$\pm$0.033 &
--26.33 & 3414 & ~ \\
J225419.23--000155.0 & 3.68 &  0.3$\pm$0.4 &  0.063  &   --0.010$\pm$0.050 &
--26.73 & 5650 & ~ \\
J225759.67+001645.7 & 3.75 & 0.4$\pm$0.4 & 0.086   &  --0.022$\pm$0.10
& --27.19 & 5520 & ~ \\
J230952.29--003138.9 & 3.95 &  --0.2$\pm$0.5 &  0.058  &
0.088$\pm$0.044 & --26.82 & 3925 & ~ \\
J235718.35+004350.4 & 4.34 & 1.8$\pm$0.6$^*$ & 0.041 & 0.110$\pm$0.027$^*$ &
--26.59 & 2650 & ~ \\
\hline
\end{tabular}
~$^c3\sigma$ limits at 5 GHz (see Fan et al. 1999 for a detailed 
discussion of J1532--0039).
\end{table}

\clearpage
\newpage

\begin{figure}
\psfig{figure=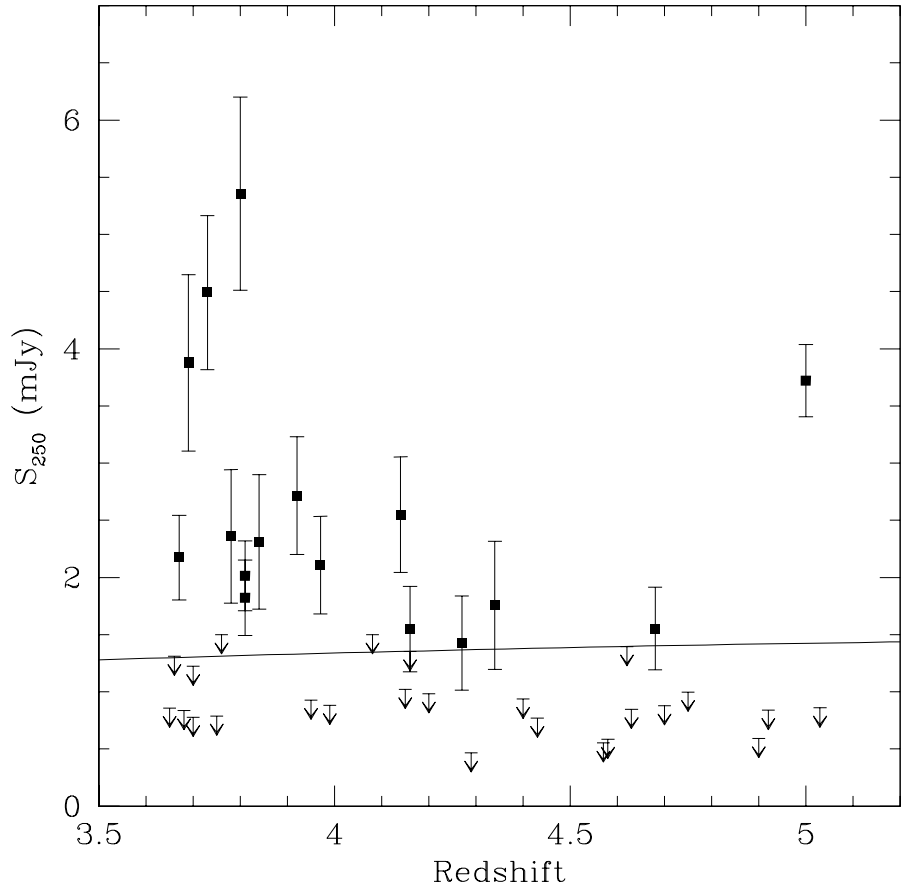,width=5in}
\caption{The relationship between 
redshift and S$_{250}$ for the SDSS high redshift QSOs listed in Table
1. The squares with error bars indicate sources detected at $>
2\sigma$. The arrows represent $2\sigma$ upper limits for
non-detected sources. 
The solid line is the expected flux density of the
starburst galaxy Arp 220 as a function of redshift.
}
\end{figure}

\clearpage
\newpage

\begin{figure}
\psfig{figure=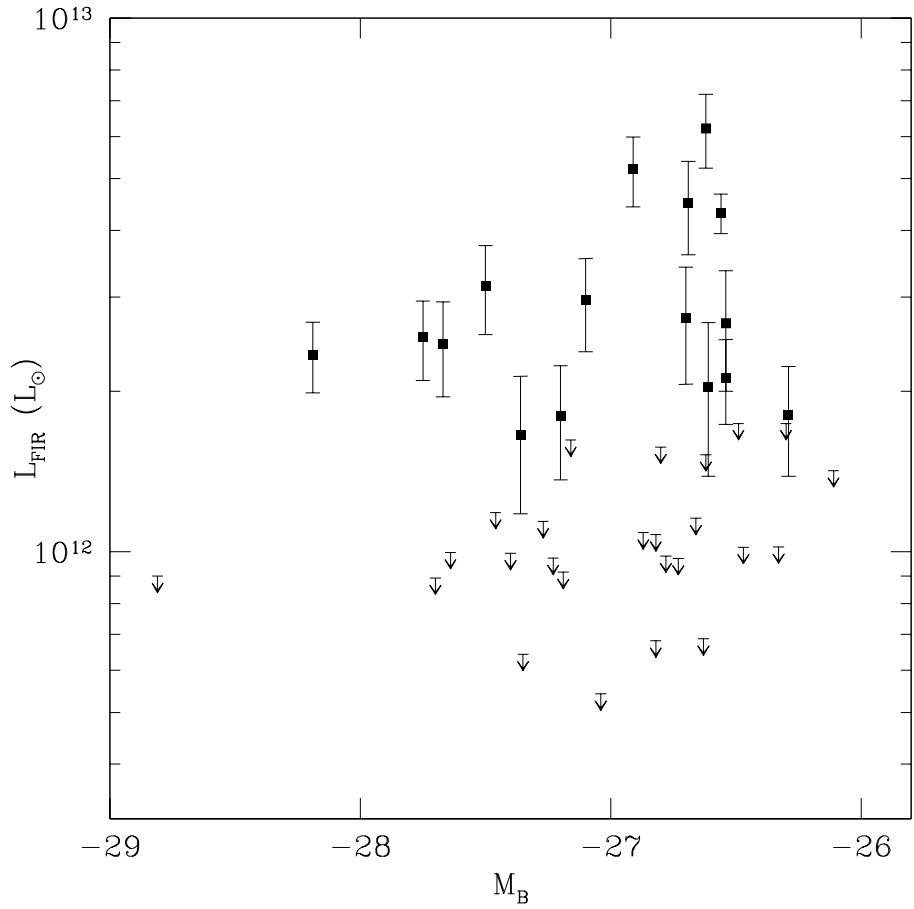,width=5in}
\caption{The relationship between M$_{\rm B}$ and 
L$_{\rm FIR}$ for the
SDSS high redshift QSOs listed in Table 1. 
The symbols are the same as in Figure 1.
}
\end{figure}

\clearpage
\newpage

\begin{figure}
\psfig{figure=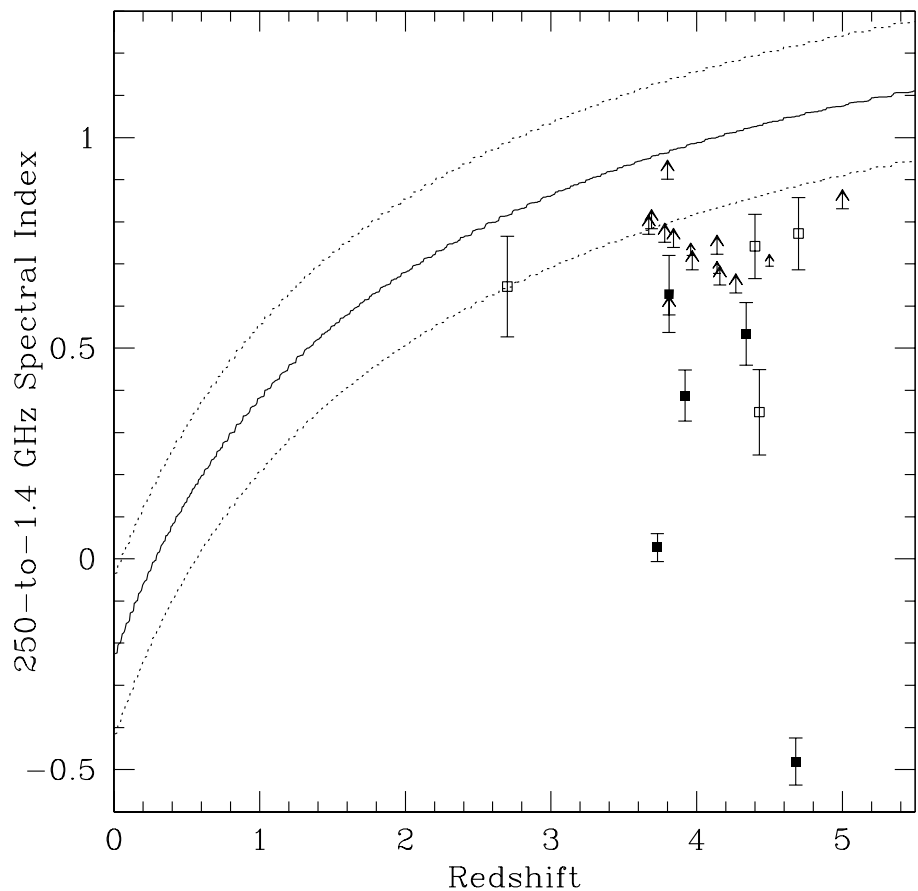,width=5in}
\caption{The solid curve shows the relationship between 
redshift and the observed spectral index between 1.4 and 250 GHz
for star forming galaxies, as derived from the models presented in 
Carilli \& Yun (2000). The dotted lines show the rms scatter in the
distribution. The solid symbols show data for SDSS high redshift QSOs
that were detected at 250 GHz. The open symbols are for APM QSOs
detected at 250 GHz (Omont et al. 1996a). The squares are for sources
detected at 1.4 GHz, while the arrows show lower limits (2$\sigma$)
to the spectral indices for sources that were not detected at 1.4
GHz. Smaller arrows are for the APM QSOs.
}
\end{figure}

\clearpage
\newpage

\begin{figure}
\psfig{figure=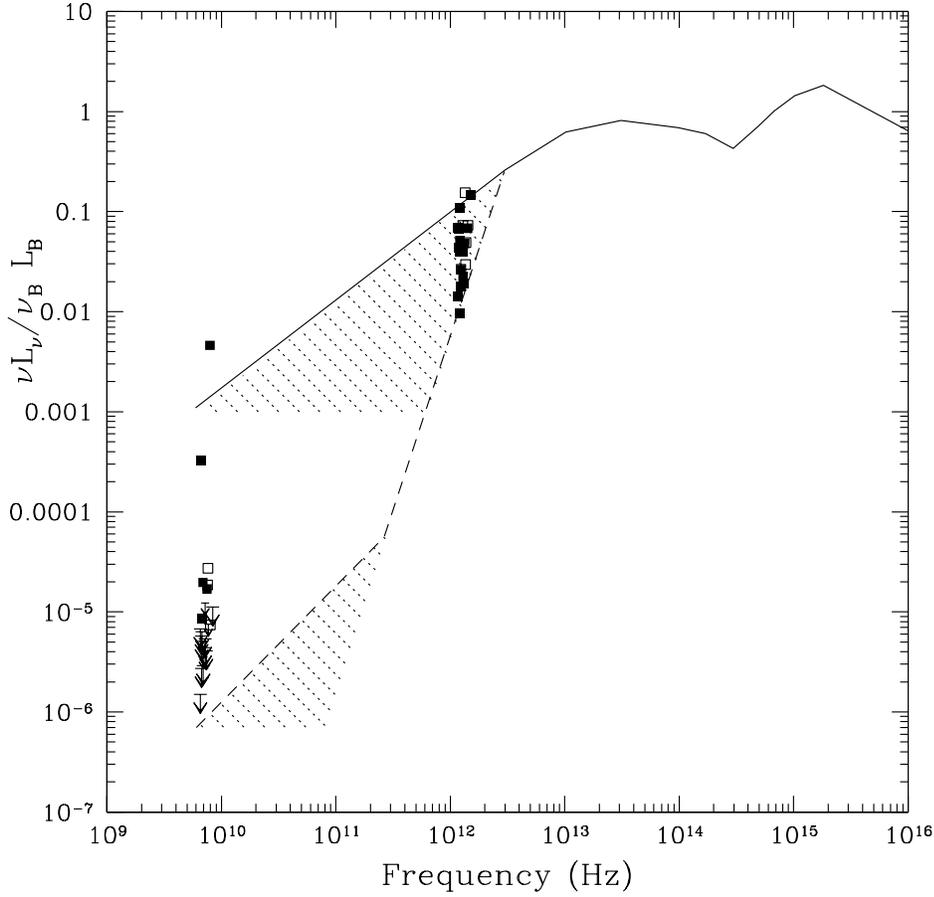,width=5in}
\caption{The curves show the
radio through UV Spectral Energy Distributions (SEDs) for QSOs, taken
from Sanders et al. 1989.
All data have been normalized to the rest frame 
blue spectral luminosity. 
For frequencies above
$3\times10^{12}$ Hz, the solid curve shows the mean spectral energy
distribution for the PG QSO sample, which is 
approximately the same
for radio-loud and radio-quiet sources. Below
$3\times10^{12}$ Hz, the hatched regions
show the allowed ranges given the scatter in the
observed properties of the PG QSO sample at cm and mm wavelengths,
with the solid line delineating the radio-loud sources and
the dashed-line the radio quiet sources. 
The solid symbols show the values derived for the SDSS QSOs 
detected at 250 GHz as listed in
Table 1, while the open symbols are for APM QSOs
detected at 250 GHz (Omont et al. 1996a). The arrows
show (2$\sigma$) upper limits at 1.4 GHz. 
The small arrows are for the APM QSOs.
}
\end{figure}

\clearpage
\newpage

\end{document}